\documentclass[%
 reprint,
superscriptaddress,
 amsmath,amssymb,
 aps,
floatfix,
]{revtex4-2}

\usepackage{graphicx}
\usepackage{dcolumn}
\usepackage{bm}
\usepackage{lipsum}
\usepackage{float}
\usepackage{hyperref}
\hypersetup{
   colorlinks=true,
   linkcolor=blue,
   filecolor=magenta,      
   urlcolor=cyan,
   pdftitle={Overleaf Example},
   pdfpagemode=FullScreen,
   }

\begin{document}

\preprint{APS/123-QED}

\title{Cascaded injection locking of optomechanical crystal oscillators}

\author{David Alonso-Tomás}
\affiliation{MIND-IN2UB, Departament d'Enginyeria Electrónica i Biomédica, Facultat de Física, Universitat de Barcelona, Martí i Franquès 1, Barcelona 08028, Spain}%
\author{Guillermo Arregui}
\affiliation{DTU Electro, Department of Electrical and Photonics Engineering, Technical University of Denmark, Ørsteds Plads 343, Kgs. Lyngby, DK-2800, Denmark}%
\author{Laura Mercadé}
\affiliation{Nanophotonics Technology Center, Universitat Politècnica de València, Camino de Vera s/n, 46022 Valencia, Spain}%
\author{Alejandro Martínez}
\affiliation{Nanophotonics Technology Center, Universitat Politècnica de València, Camino de Vera s/n, 46022 Valencia, Spain}%
\author{Amadeu Griol}
\affiliation{Nanophotonics Technology Center, Universitat Politècnica de València, Camino de Vera s/n, 46022 Valencia, Spain}%
\author{Néstor E. Capuj}
\affiliation{Depto. F\'{i}sica, Universidad de La Laguna, 38200 San Crist\'{o}bal de La Laguna, Spain}%
\affiliation{Instituto Universitario de Materiales y Nanotecnolog\'{i}a, Universidad de La Laguna, 38071 Santa Cruz de Tenerife, Spain}%
\author{Daniel Navarro-Urrios}
\email{dnavarro@ub.edu}
\affiliation{MIND-IN2UB, Departament d'Enginyeria Electrónica i Biomédica, Facultat de Física, Universitat de Barcelona, Martí i Franquès 1, Barcelona 08028, Spain}%
%
%

\date{\today}

\begin{abstract}
Optomechanical oscillators stand out as high-performance and versatile candidates for serving as reference clocks in sequential photonic integrated circuits. Indeed, they have the unique capability of simultaneously generating mechanical tones and optical signal modulations at frequencies determined by their geometrical design. In this context, the concept of synchronization introduces a powerful means to precisely coordinate the dynamics of multiple oscillators in a controlled manner, thus increasing efficiency and preventing errors in signal processing photonic systems or communication interfaces. In this work, we demonstrate the cascaded injection locking of a pair of silicon-based optomechanical crystal cavities to an external reference signal that subtly modulates the laser driving one of the oscillators. Both cavities interact solely through a weak mechanical link, making the extension of this synchronization mechanism to an increased number of optomechanical oscillators within a common chip more feasible than relying solely on optical interactions. Thus, the combination of the obtained results, supported by a numerical model, with remote optical injection locking schemes discussed in the literature, lays the groundwork for the distribution of reference signals within large networks of processing elements in future phonon-photon hybrid circuits.
\end{abstract}

\maketitle


Among the diverse array of optomechanical (OM) devices \cite{Aspelmeyer}, optomechanical oscillators (OMOs) have attracted significant attention due to their potential in addressing various scientific and technological challenges. These oscillators are exceptional platforms, characterized by high-amplitude, self-sustained and coherent mechanical motion driven and controlled by optical fields \cite{Kippemberg}. In this context, although they have already found applications in precision sensing \cite{Yu, Pan, Liu} as well as in frequency generation and conversion \cite{Hossein, Mercade, Mercade2, Ghorbel} among others, their properties make them excellent candidates for serving as clock signals in photonic integrated processing circuits. In fact, with the recent advances in this field and the possibility of combining optomechanical crystal cavities (OMCs) with routed microwave phonons \cite{Painter}, the approach could be naturally extended to chip-integrated phonon-photon platforms where multiple of these oscillators interact with each other. However, the extension of this system to a large network of OMOs generates the need of a mechanism for preventing errors related to frequency dispersion and improving the efficiency of individual OMOs in terms of phase noise and frequency jitter. 
\\
In this context, injection locking, an established process in electronic and radio-frequency (RF) systems pioneered by the seminal paper of Adler et al. \cite{Adler}, holds the key to synchronize actions and further improve the overall response of the system. In an injection locking mechanism, the frequency and phase of an oscillator become synchronized with those of an external signal. This external signal is injected into the oscillator and originates from another local oscillator (LO) that acts as a reference clock. The first observation of injection locking in an OMO was documented in a silicon microtoroid subjected to a modulation of the same optical drive that leads to the optomechanical oscillations \cite{Hossein2}. Subsequently, additional works have demonstrated injection locking in various OMOs using a similar optical injection scheme where the LO comes from a signal generator \cite{Luan, Shlomi, Arregui} or even from another OMO \cite{Shah, Li, David}. Injection locking has also been achieved through electrical capacitive actuation \cite{Pitanti, Bekker} and by mechanical actuation with propagating acoustic waves \cite{Huang}.
\\
In order to distribute the reference signal within a chip, a natural strategy is that of synchronizing several OMOs to the external signal, physically placed at different locations of the chip, in a cascaded injection locking (CIL) configuration. To achieve this, supplying a modulated optical pump concurrently or sequentially to multiple OMOs, either in a parallel or series arrangement, appears to be a viable approach. However, the practical execution of this strategy on a reasonably large scale presents a significant challenge due to the requirement that all these OMOs must share an identical resonant optical wavelength. In this context a small number of reports have claimed the spontaneous synchronization of OMOs sharing a common optical mode \cite{Bagheri, Zhang, Zhang2}.  In fact, up to date there is only one work reporting the cascaded locking of individual OMOs \cite{Santos}. In that study, the OMOs are optically interconnected by means of a common waveguide in a sequential arrangement and no external reference was fed to the first OMO. Even though this configuration allows scalability, its implementation with silicon microdisk difficult the direct extraction of coherent mechanical signals. Moreover, OMCs suffer from stronger optical dispersion, making an all-optically mediated synchronization complicated. \\

Here, we demonstrate an approach for achieving CIL of the dynamics of a pair of chip-integrated OMCs acting as OMOs to an external reference signal. In our platform, the resonators are optically isolated from each other and interact through a weak mechanical link. Exploiting a mechanical interaction in spite of an optical one allows a more straightforward means of achieving spontaneous synchronization of OMCs \cite{Colombano}, given that the mechanical resonant frequency dispersion among different OMCs is usually much smaller than the optical counterpart. The interaction is considered unidirectional since one of them (referred to as the "main" or "leader") oscillates with significantly larger amplitude and injection-locks to the reference tone, while the other one (designated as the "secondary" or "follower") spontaneously synchronizes to the main one and/or the external reference only through the weak mechanical interaction. In this context, we also report that it is possible to injection-lock the secondary OMO to the external reference, even though the main OMO is the one receiving the reference tone and oscillating independently. 
\begin{figure*}[t]
    \centering    
    \includegraphics[width =\linewidth]{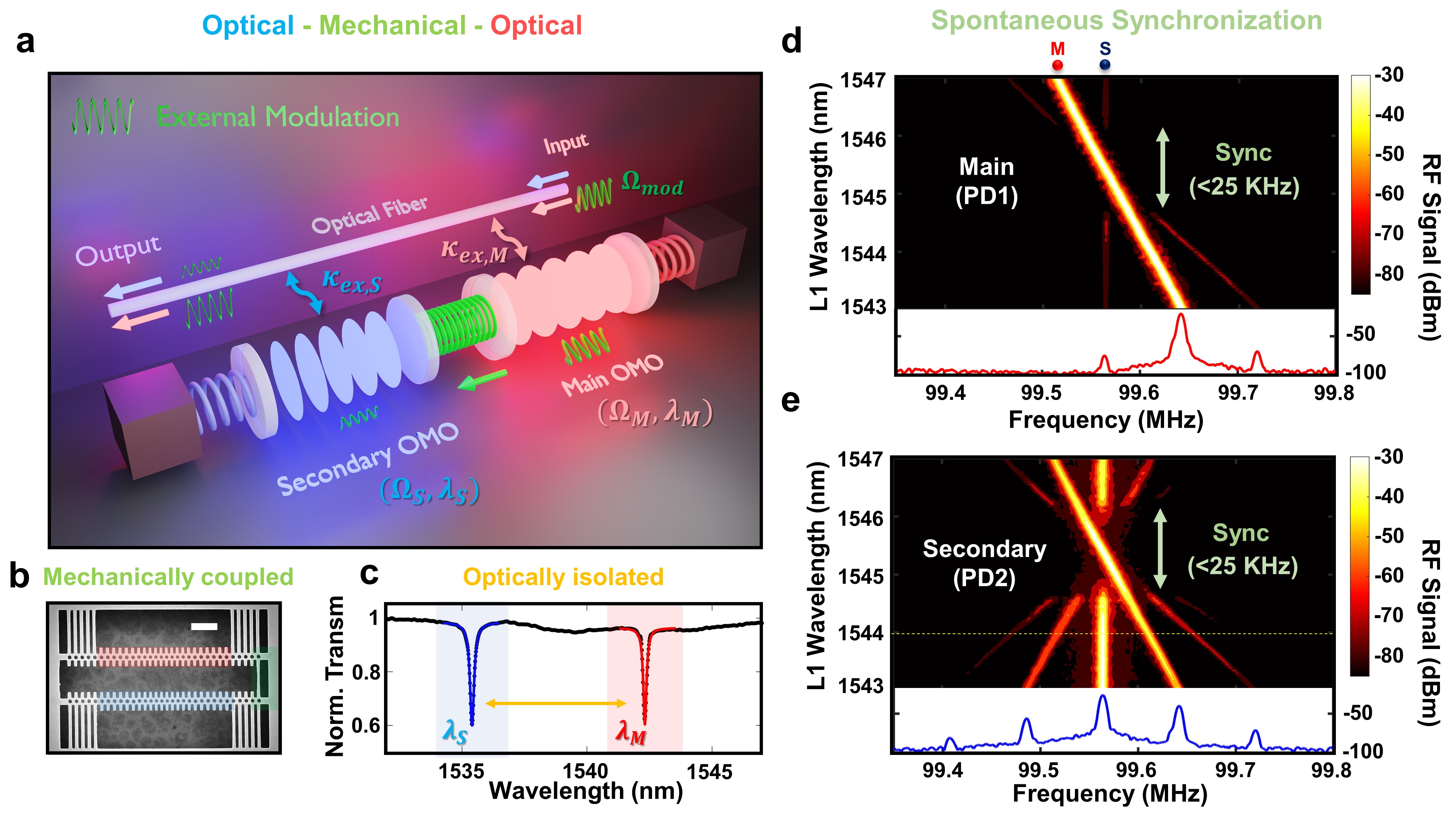}
    \caption{Studied configuration and spontaneous synchronization of two optomechanical crystal oscillators (OMOs) \textbf{a} Conceptual illustration of the proposed system, where both OMOs, represented as two Fabry-Perot cavities with one moveable mirror attached to a spring, interact through another spring of lower elastic constant. Both structures support optical and mechanical modes with wavelengths and mechanical angular frequencies $\lambda_{i}$ and $\Omega_{i}$, where i = M,S denotes main and secondary respectively. Laser light propagating through a tapered fiber is evanescently coupled to each optical mode with an extrinsic rate $\kappa_{ex,i}$. An external modulation with frequency $\Omega_{mod}/2\pi$ is applied to the light in resonance with the main OMO, which is transduced in the secondary channel due to the weak interaction between both cavities. \textbf{b} SEM micrograph of two close-by silicon optomechanical crystal nanobeams that are connected by a mechanical link. White scale bar is 2 $\mu m$. Both beams support optical cavity modes that are spectrally and spatially isolated, as shown in the optical transmission spectrum of panel \textbf{c} (optical resonance at room temperature appear at $\lambda_{S,0} = 1535.4$nm and $\lambda_{M,0} = 1542.4$nm). \textbf{d} and \textbf{e} Radio-frequency spectra of the optical transmission associated to the main and the secondary OMOs respectively, for different values of the wavelength of the laser driving the main OMO ($\lambda_M$). The spectral range in which spontaneous synchronization occurs is highlighted with green dashed lines. The wavelength of the laser driving the secondary is fixed at $\lambda_S = 1537 nm$. A yellow dashed line in panel illustrates the conditions that are kept in the experiments with an external modulation.}
    \label{fig: setup}
\end{figure*}

\section*{Results}
\textbf{Optical-Mechanical-Optical Configuration}. The experiment conducted in this work closely parallels the scenario depicted in Figure \ref{fig: setup}a. Two OM cavities, represented as Fabry-Perot cavities with a mirror attached to a spring, weakly interacts through another spring of lower elastic constant.  Both cavities support different optical resonances, with red representing the main OMO ($\lambda_M$) and blue representing the secondary one ($\lambda_S$), which are resonantly pumped by evanescently coupled laser light propagating through a tapered fiber. Each oscillator has a resonant mechanical angular frequency of $\Omega_i$, where $i = M, S$ denotes main and secondary respectively. The oscillation in the cavity length leads to a spectral shift of the optical resonances. The strength with which these resonant frequencies ($\omega = 2\pi c/\lambda$) are pulled (linearly) in terms of mechanical displacement can be defined as $G = \partial \omega/\partial x$, which is related to the vacuum optomechanical coupling rate $g_0 =  G x_{\mathrm{ZPF}}$. Here, $x_{\mathrm{ZPF}} = \sqrt{\hbar/2m\Omega}$ represents the mechanical zero-point fluctuation of the mechanical mode with effective mass m and frequency $\Omega$. It is worth noting that if the incident laser wavelength is in resonance with an optical mode, the light transmitted will undergo a modulation enveloped by the characteristic Lorentzian shape of a Fabry-perot cavity mode. Hence, input light acts as a probe to observe the mechanical oscillation of OMOs, but at the same time can play an active role in the optomechanical system through radiation pressure forces.\\

Both OM cavities are optically driven into a state of high amplitude, coherent and self-sustained mechanical motion, referred to as mechanical lasing, in such a way that one oscillates with significantly larger amplitude than the other. This configuration effectively emulates a leader-follower arrangement of oscillators \cite{Pikovski}. Subsequently, a periodic external force is applied solely to the main oscillator, resulting from an external modulation in the amplitude of the input light in resonance with its supported optical mode. Thus, the secondary one perceive its actuation only through the weak mechanical interaction between the systems. If the frequency of the external modulation is close enough to the resonant frequency of one of the OMOs, its dynamics can be locked to the external modulation \cite{Arregui}. CIL occurs when the two resonators synchronize to the external perturbation.\\

Consider the presented scheme, where the interaction can be understood as a restoring force emerging from the mechanical link, the equations governing the dynamics of the leader-follower arrangement of oscillators can be described as two reactively coupled linear forced harmonic oscillators:
\begin{equation}
\ddot{x}_i + \Gamma_{i} \dot{x}_i + \Omega_{i}^2 x_i  = \frac{F_{0,i}(t)}{m_{i}} - 2J\Omega_{i}x_M\delta_{i,S}\\
\label{eq: mec}
\end{equation}
where $\delta$ is the dirac delta, J corresponds to the reactive coupling strength, $\Gamma_{i}$ is the mechanical decay rate of each resonator and $F_{0,i}(t) = g_{0,i} \hbar n_{0,i}(t)/x_{ZPF}$, the radiation pressure force exerted by the temporally dependent intra-cavity photon number ($n_{0,i}$). The temporal behaviour of this magnitude is governed by the self-induced modulation associated to the driving mechanism, as well as a periodic external modulation with frequency ($f_{mod} = \Omega_{mod}/2\pi$) in the case of the main OMO equation (see \cite{Arregui, Dani2} and supplementary section S3 for more details). The coupling term only appears in the equation corresponding to the follower oscillator since the main cavity oscillates with much higher amplitude than the secondary one, emulating the unidirectionality of the system.\\

\textbf{Tested Device and Experimental Setup}. In this work we use silicon one-dimensional OMCs, behaving as OMOs. These structures are planarly chip-integrated suspended nanobeams which behave as photonic crystals with a defect region, whose design and fabrication methodology are detailed in the Methods section. The tested device is composed of a pair of nominally identical OMOs, where the outermost five cells of each nanobeam are anchored to the frame, restricting the flexural modes of the geometry to its central area. The specific geometry of the OMO devices, allows positioning the tapered fiber between both geometries, enabling their simultaneous optical excitation. Here, the OM coupling is understood as the change in resonant optical frequency induced by the moving boundaries and photo-elastic effect contributions of the mechanical mode \cite{Eichenfield}. 

The interaction between both OMOs is provided through an engineered mechanical link (Fig. \ref{fig: setup}b) between them, placed on the anchored zone, thus ensuring a weak mechanical coupling. In this platform, the pumping mechanism enabling mechanical lasing of the OMOs is the anharmonic radiation pressure modulation $F_{0,i}(t)$ due to the activation of a thermo-optic/free-carrier dispersion self-pulsing (SP) limit-cycle that emerges in silicon under a certain threshold of number of photons inside the cavity (see Methods and supplementary information S2). This driving mechanism is useful when the system is in the sideband unresolved regime, where the decay rate of the optical cavity modes is much higher than the frequencies of mechanical resonators, and has been reported by us and other groups previously \cite{Johnson, Dani, Luiz}. 
In particular, we have driven to the mechanical lasing regime the three-antinodes flexural mechanical modes of the secondary and main OMOs, whose room temperature natural frequencies are ($f_S$,$f_M$) = ($\Omega_{S}$, $\Omega_{M}$)/2$\pi$ = (99.57, 99.65) MHz. Here, fabrication disorders break the symmetry, causing the OMCs to exhibit different mechanical natural frequencies. These fabrication deviations also lead to an increase in the OM coupling of these in-plane flexural modes to hundreds of kHz \cite{Colombano}, even though the simulated OM coupling of the nominal geometry is null. \\

The experimental setup, which is schematically represented in the Methods section, implements the configuration described in Fig. \ref{fig: setup}a. The two OMOs are excited using tunable diode lasers, respectively denoted by L1 and L2, whose polarization state is set to be transverse electric (i.e., electric field in the plane of the sample), using fiber polarization controlers (FPC). Light emitted from L1 is amplitude modulated using a Mach-Zehnder electro-optic modulator (EOM), with a half-wave voltage $V_\pi$ = 3.5 V. A signal generator (SG) provides the periodic external modulation. Both lasers are then combined into a single tapered fiber that has been shaped into a microloop at its thinnest region \cite{Fabero}. The bottom part of the loop acts as a probe that allows the local excitation of the optical cavity modes of the OMOs if the cavity regions are in the near field of the fiber and the lasers are in resonance with the respective optical modes. The signal decoupled from the resonators is then split and passed through Fabry-Perot wavelength filters (WF), to afterwards impinge on two photodetectors (PD1 and PD2). The electrical output can be analyzed in the frequency domain by a spectrum analyzer (SA) and in the temporal domain with a 4-channel oscilloscope (OSC). All the electronic equipment owns a response bandwidth of 11 GHz. 

The optical transmission spectra of the OMOs displayed in Figure \ref{fig: setup}c shows that the first optical resonances of each of the OMOs are separated by about 7nm (1535.4 nm for the secondary OMO and 1542.4 nm for the main one). This large separation (0.45\%) emerges from fabrication imperfections and the presence of the fiber, since optical properties are very sensible to small deviations in the geometry or the surrouding media. In this case, the optical power coupled fraction for both OMOs is roughly 35\% and the loaded optical quality factors are $Q_S=7.3 \cdot 10^3$ and $Q_M=11.3 \cdot 10^3$, while the intrinsic values are $9.4 \cdot 10^3$ and $14.3 \cdot 10^3$, respectively. This configuration ensures the absence of direct optical interaction between the two OMOs.  

\textbf{Spontaneous Synchronization}. First, we focus on the case where no external modulation is provided. As previously explored in \cite{Colombano}, the weak interaction between the two self-sustained OMOs in the mechanical lasing regime can induce spontaneous synchronization of the follower oscillator to the leader within a specific range of detuning between their natural frequencies. In the present study, the oscillation frequencies of both OMOs just above the self-oscillation threshold are too far apart to enable spontaneous synchronization of their dynamics. To control the frequency difference, we increased the wavelength of the laser driving the OMO with the larger mechanical frequency, which in this case is the main cavity. This leads to an increase in the time-averaged number of intracavity photons, causing more photons to be absorbed and the geometry to heat up. Figure \ref{fig: setup}d and \ref{fig: setup}e illustrate this experiment. The top panels show, in a contour plot representation, the RF spectrum of the signal detected by PD1 and PD2 as the wavelength of L1 is modified. The bottom panels illustrate the RF spectrum at the initial wavelength. The spectra corresponding to the main OMO (Figure \ref{fig: setup}d) show that $f_M$ reduces gradually as the wavelength increases, in an almost linear way due to a relaxation of the elastic constants of the material \cite{Navarro}. Here, the only signs of the mechanical feedback coming from the secondary OMO appear as RF peaks at $f_S$ and $2f_M - f_S$. These sidebands slightly approach $f_S$ and disappear within a range of wavelengths between 1544.8 nm and 1546.1 nm, which corresponds to an OMO frequency difference of $|f_M - f_S| <$ 25 KHz. \\

The secondary OMO dynamics displayed in Figure \ref{fig: setup}e reveals a similar behaviour. At the initial condition, a high-amplitude RF peak is placed at $f_S$, surrounded by two principal sidebands at $f_M$ and $2f_S - f_M$ of much larger amplitude than in the previous case, which confirms the unidirectional character of the interaction. Here, it is clear that the range in which the sidebands disappear matches with an abrupt jump of $f_S$ becoming now equal to $f_M$, which is evidence of spontaneous synchronization of the secondary OMO dynamics to that of the main OMO. The magnitude of the main RF peak associated to the secondary OMO oscillation keeps the same value along the whole spectral range analyzed in the experiment, which rules out the possibility of resonant forcing. The wide range of tunability of $f_M$, expanded when compared to the case reported in Ref. \cite{Colombano}, allows displaying the whole synchronization range, i.e., the whole Arnold tongue. Under these conditions, the Arnold tongue appears to be symmetric with respect to the natural value of $f_S$. 

 \begin{figure*}[t!]
    \centering    
    \includegraphics[width =\linewidth]{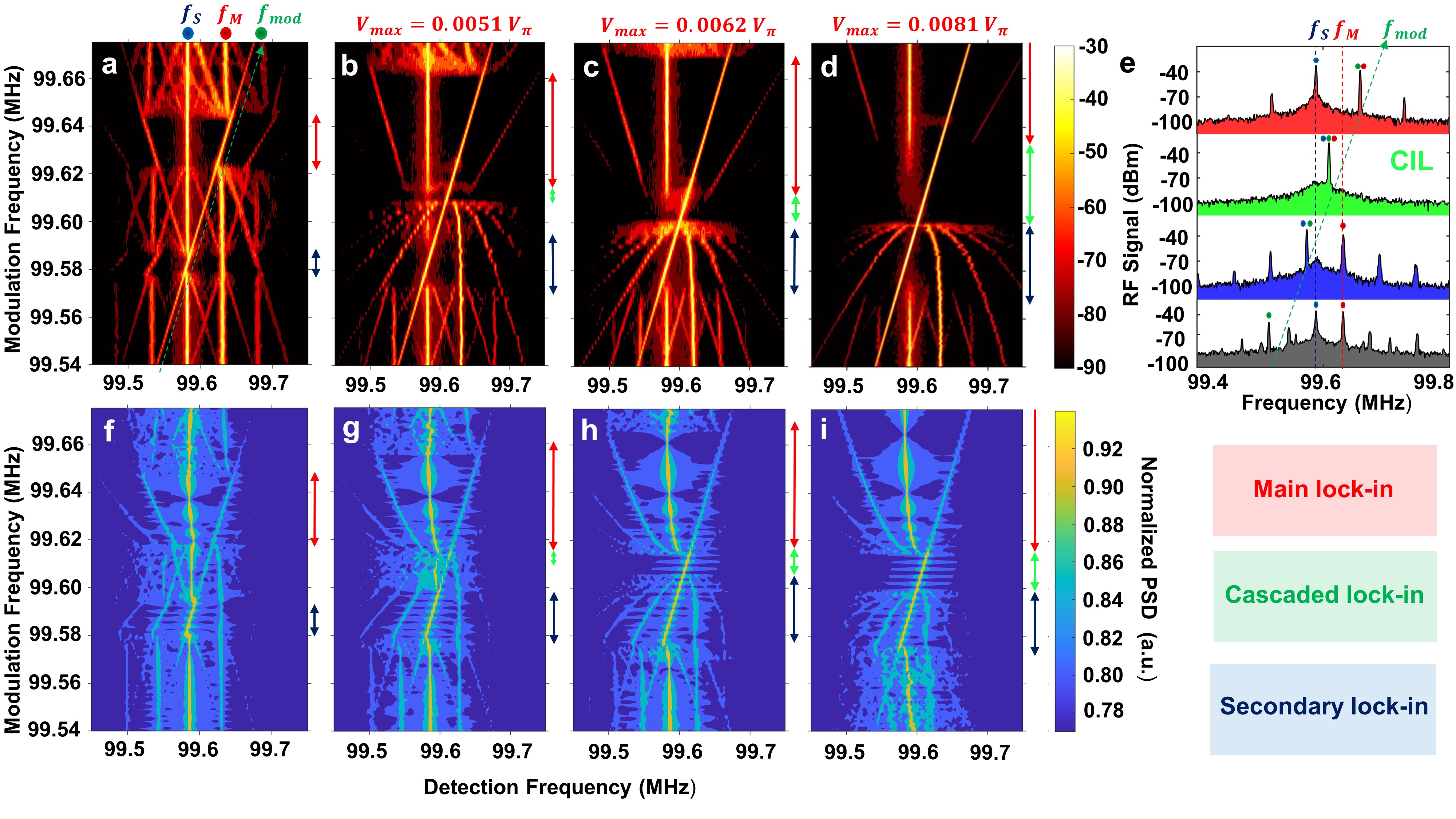}
    \caption{Colormaps of the power spectral density (PSD) of the optical transmission associated to the secondary OMO when an external harmonic tone modulates the main one. \textbf{a} Explanation of the observed RF peaks at low amplitude of external modulation. The vertical axis is associated to the modulation frequency ($f_{\mathrm{mod}}$) and is highlighted with a green dashed arrow. The RF peaks associated to the natural oscillation of the secondary resonator and the main one are highlighted with blue and red dots. The rest of the peaks are sidebands appearing at different frequency combinations of the mentioned peaks. The amplitude of the voltage applied to the modulator ($V_{\mathrm{max}}$) is 0.0042$V_\pi$. \textbf{b-d} Set of panels representing PSD versus $f_{\mathrm{mod}}$ for $V_{\mathrm{max}}$ varying from $V_{max}$= 0.0051$V_\pi$ to $V_{\mathrm{max}}$ = 0.0081$V_\pi$. The range of $f_{mod}$ in which secondary OMO, main OMO, and cascaded injection locking are observed are highlighted with blue, red and light green arrows. \textbf{e} Representative RF spectra of each regime observed in the case of higher amplitude of modulation ($V_{\mathrm{max}}$ = 0.0081$V_\pi$). Grey, blue, green and red areas denote free-running, secondary, cascaded and main lock-in respectively. These colors are used also to highlight the position of the RF external tone, and the main and secondary oscillating frequencies (green, red and blue, respectively). \textbf{f-i} Numerical simulations showcasing similar regimes to the ones observed in experimental measurements. Selected voltage ($V_{\mathrm{max}}$) is (0.0127, 0.0196, 0.0254, 0.350) $V_\pi$, respectively.  }
    \label{fig2}
\end{figure*}

\textbf{Cascaded Injection Locking}. The previous experiment demonstrates the spontaneous synchronization of the two OMOs in a leader-follower scheme. In the upcoming studies, we keep the wavelength of L1 at 1544 nm (see the yellow dashed line in Figure \ref{fig: setup}e), where $|f_M - f_S| =$ 45 KHz, placing the two OMOs outside the Arnold tongue, i.e., they do not synchronize spontaneously. We now introduce a harmonic signal of frequency $f_{\mathrm{mod}}$ and amplitude $V_{\mathrm{max}}$ generated by the SG, which modulates the amplitude of the output of L1. Figure \ref{fig2} shows the result of monitoring the RF signal of the secondary OMO optical channel as a function of $f_{\mathrm{mod}}$ for different $V_{\mathrm{max}}$ values. When $f_{\mathrm{mod}}$ is sufficiently detuned from the natural oscillation frequencies, the RF spectrum consists of the peaks reported in Figure \ref{fig: setup}e, with the main peak at $f_S$ surrounded by two main sidebands at $f_M$ and $2f_M - f_S$ (see Figure \ref{fig2}a or the grey spectra in Figure \ref{fig2}e). Additionally, there is the contribution of the modulation tone (green dashed arrow in Figure \ref{fig2}a). Given that the modulation is applied to L1, the appearance of this tone in the signal corresponding to the secondary OMO is a consequence of the interaction with the main OMO. Indeed, it follows the spectral response of the flexural in-plane mechanical modes, which have intrinsic natural linewidths of typically 0.2 MHz at atmospheric pressure. \\

Now, as $f_\mathrm{mod}$ increases and approaches to $f_S$ we observe sidebands corresponding to the beat between the external modulation and higher harmonics of the secondary OMO signal. When the frequency of the external modulation is close enough to $f_S$, the latter suffer a frequency pulling effect towards $f_\mathrm{mod}$ which ends up with a narrow injection locking region of the secondary OMO dynamics to the external tone. By further increasing $f_\mathrm{mod}$, so that we analyze the region around $f_M$, the same effects are displayed but in a stronger extent. This is consistent with the fact that the main OMO is directly experiencing the modulation of the radiation pressure force while the secondary OMO senses a much weaker mechanical perturbation through the mechanical link. It is worth noting that the frequency pulling effect in both injection locking ranges is the expected behaviour in a phase locking mechanism \cite {Arregui}.\\

If we focus now on panels \ref{fig2}b-d, we observe that the two locking regions described before, denoted with blue and red arrows for the secondary and main OMO injection locking ranges respectively, widen as $V_{\mathrm{max}}$ increases. Above $V_{\mathrm{max}}$ = 0.0051$V_\pi$, CIL of the two oscillators to the external drive is achieved, which is highlighted with a light green arrow. The CIL range starts on the lower-frequency end of the main OMO injection locking range, which is where the frequencies of the two OMOs are closer.  Within this regime, the RF spectrum collapses into a single tone at $f_\mathrm{mod}$ (see green spectra of Fig. \ref{fig2}e). The frequency range of CIL widens as $V_{\mathrm{max}}$ increases, achieving few tens of kHz in the extreme case (see Fig. \ref{fig2}d). For large enough $f_{mod}$ values, the two OMO frequencies become too far apart to keep the CIL and it is lost, however, the main OMO remains locked to the external reference signal through a larger bandwidth. In this later situation, the secondary cavity oscillates independently again (red spectra of figure \ref{fig2}e) and sidebands corresponding to the beat between the external modulation and secondary OMO harmonics appear at a combination of their frequencies. It is important to mention that figure \ref{fig2}b and \ref{fig2}c-d represent different accesses to the CIL regime. In the first case, the secondary oscillator spontaneously synchronizes to the perturbation generated by the main one when it is locked to the external frequency. This occurs because the detuning between mechanical frequencies after the injection-locking of the main OMO, falls below the threshold for spontaneous synchronization (Fig. \ref{fig: setup}e). On the other hand, when the external modulation generates a mechanical perturbation large enough to maintain the secondary OMO locked until the main Arnold tongue, the CIL is accessed by the locking of both oscillators to the external feedback. \\

\begin{figure*}[t]
    \centering    
    \includegraphics[width =\linewidth]{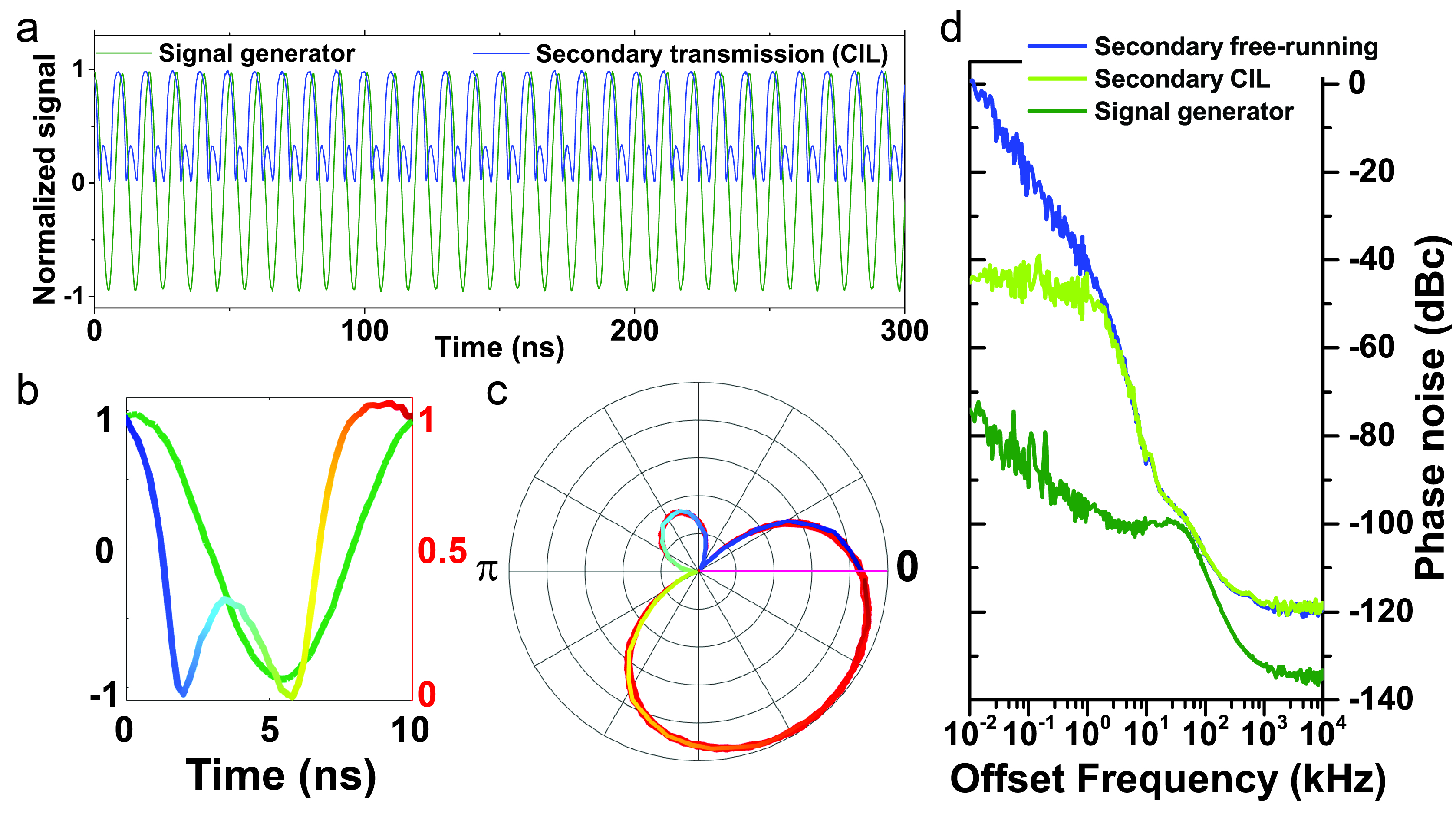}
    \caption{Temporal traces, Poincaré map and phase noise of the cascaded injection locking state. \textbf{a} Temporal traces of the reference tone (dark green) and of the optical transmission of the secondary OMO (blue) as recorded simultaneously in two channels of the oscilloscope. \textbf{b} Enlargement of a single period of the two signals. \textbf{c} Time trace of the secondary OMO transmission signal in a polar representation, where the distance to the centre is the transmission amplitude value and the angle is the phase difference with respect to the maximum of the modulation signal. The total temporal register used for this representation is 1 $\mu s$. d Phase noise measurements of the free-running secondary OMO (blue), radio-frequency reference tone (dark green) and of the  secondary (light green) in the CIL regime. }
    \label{fig4}
\end{figure*}

The system of coupled differential equations corresponding to the mechanical resonators (Eq. \ref{eq: mec}) and the SP mechanism (see Methods and supplementary S2) can be solved numerically to compute the temporal dynamics of intra-cavity photons and, hence, the transmission temporal traces for the main and secondary OMO (see Supplementary Material S3). In order to compare numerical simulations with experimental data, a Fast Fourier Transform (FFT) is performed on the computed transmission temporal trace of the secondary OMO as $f_\mathrm{mod}$ is swept over a range between the natural frequencies. Figures \ref{fig2}g-i show the same contour RF representation used to visualize experimental measurements for different amplitudes of the external modulation. Here, simulations clearly reproduce the different regimes observed, including the injection locking range induced in the secondary OMO by the mechanical perturbation, as well as the CIL of both OMOs for higher amplitudes of modulation.\\

\textbf{Temporal Traces and Phase Noise}. Temporal measurements of the dynamics of the two OMOs in the CIL regime ($f_\mathrm{mod}$ = 99.61 MHz) were performed using the oscilloscope. Figure \ref{fig4}a represents the time traces of the external modulation signal and the secondary OMO when it is locked to the external tone by means of CIL. Both traces are taken simultaneously using the first one as a trigger signal in the oscilloscope. A zoom of a single period of the two traces is shown in Figure \ref{fig4}b, where the secondary OMO transmission is now represented in a color scale. Figure \ref{fig4}c displays the transmission signal in a polar representation suitable to avoid self-intersections and to superimpose all the full cycles recorded in the transmission temporal trace, which spans a total of 1 $\mu s$, i.e., about 100 cycles. This is a specific type of Poincaré map in which the transmission of the secondary OMO is sampled at a frequency equal to $f_\mathrm{mod}$ over all the cycles. The radius is associated to the limit-cycle optical transmission trace and the angle to the relative phase of the harmonic modulation signal with respect to its maximum, i.e., we express the reference signal as $\cos(2 \pi f_\mathrm{mod}t)$, where $2 \pi f_\mathrm{mod}t$ is the polar angle of the plot. To better illustrate this, the transmission curves have been plotted in both sub-panels with a common color scale linked to each phase point. If there is no frequency locking between the two signals, the resulting curve will slightly rotate with time, filling the whole polar space if the temporal acquisition is long enough. However, the trajectory drawn in Figure \ref{fig4}c repeats itself in every cycle and no drift is observed regardless of the acquisition time.\\

Finally, high-precision linewidth measurements were registered using a phase noise analyzer integrated in the SA. In Figure \ref{fig4}d we compare the phase noise of the free-running secondary OMO (blue), with that of the signal generator (dark green) and that of the secondary OMO (light green) within the CIL regime. The part of the phase noise associated to the secondary OMO that is greatly improved is that involving the long-term stability of the OMOs, that is, below offset frequencies of some kHz. We associated this effect to the weak nature of the synchronization mechanism. Indeed, in Ref \cite{Colombano} we demonstrated that it takes several hundreds of oscillations to restore the spontaneous synchronization dynamics when the system is perturbed externally. The performance of the secondary OMO remains the same for frequencies above 10 kHz, which involve faster noise mechanisms \cite{Rokhsari} to which spontaneous synchronization does not provide any improvement.

\section*{Discussion}
We have studied the dynamics of a pair of Si-based OMCs acting as OMOs, that are weakly coupled mechanically. In the absence of an external signal, the system behaves as a leader-follower configuration, where the main resonator exhibits a larger oscillation amplitude. Additionally, we observed regions of injection-locking in both OMO dynamics when introducing an external harmonic modulation on the laser driving the main oscillator at frequencies close to their natural ones. This behavior is anticipated in the case of the main OMO, given that the modulation is directly applied to it \cite{Arregui}, but it is not straightforward to observe in the case of the secondary OMO. The origin of this latter injection-locking mechanism lies in the mechanical interaction between both integrated OMOs, which remains effective even if the main one is not injection-locked to the reference. As a result of the study, we have unequivocally demonstrated, both in the frequency and temporal domains, the cascaded injection-locking of the OMOs to an external reference signal, whereby the main OMO is injection-locked to the reference and the secondary OMO is spontaneously synchronized to the mechanical perturbation coming from the link. All these results were qualitatively reproduced with numerical simulations modeling the non-linear dynamics of the coupled system of oscillators. \\

The reported cascaded injection-locking mechanism in chip-integrated Si-based OMOs, where the interaction is provided through a mechanical link, offers an alternative for distributing signals among OMOs that do not share a common optical mode. This scheme enables the use of multiple optical channels and avoids issues related to optical frequency dispersion among nominally equivalent chip-integrated OMCs due to fabrication disorder. In this context, these geometries not only allow for the extraction of coherent mechanical signals through the substrate, leading to photon-phonon hybrid circuits \cite{Painter}, but also can be optically excited by common on-chip bus waveguides \cite{ACS}. Hence, the proposed platform could be extended to multiple oscillators sequentially locked to the one receiving a reference signal, with the adaptation time of the OMOs to the external reference being an important factor to consider. Combining the presented configuration with an optical injection locking scheme \cite{David} opens a path for distributing reference signals along distant clusters of processing elements interacting remotely. The complex nonlinear dynamics emerging in this type of platform could be leveraged not only for signal distribution but also for advanced applications, including enhanced frequency generation or conversion, as well as ultrasensitive measurements by monitoring the collective dynamics of the system \cite{Lamberti}.\\

Overall, this study represents a necessary step towards achieving full dynamic control of phonon lasers through synchronization, which is a critical element in the development of phonon-photon hybrid circuits.

\section*{Methods}
\textbf{OM crystal design}. The experimental structures employed in this study consist of one-dimensional silicon-based OM crystal cavities. These are standalone nanobeams simultaneously functioning as photonic and phononic crystals. The unit cell features a central mass with a pitch (a) of 500 nm, a central hole with a radius (r) of 150 nm, and stubs extending from the top and bottom of the central mass with a length (d) of 250 nm (see Figure \ref{fig: design}). The incorporation of holes and stubs in the same geometry enables independent modification of the photonic or phononic bandgap \cite{Gomis}. Parameters are specifically design so that the geometry acts as a full photonic and phononic mirror in the frequency ranges of 200 THz and 4 GHz. The adiabatic reduction of these parameters at the center of the crystal, to $\Gamma$ = 85\% of their original values, forms a defect region giving rise to confined modes (which are not used in this work). The OM crystal comprises 32 unit cells, collectively spanning a length of approximately 16 $\mu$m.\\

\textbf{OM crystal fabrication}. In this work, the studied device comprises two equivalent one-dimensional OM crystals joined by one of its stubs at one end. These crystal pairs were fabricated on conventional silicon-on-insulator (SOI) SOITEC wafers, featuring a silicon layer with a thickness of 220 nm, a resistivity of approximately 1 to 10 $\Omega$ cm$^{-1}$, and p-doping at around $10^{15}$ cm$^{-3}$. The SOI wafer included a buried oxide layer with a thickness of 2 $\mu$m. The design was etched onto a poly-methyl-methacrylate (PMMA) resist film with a thickness of 100 nm using electron beam technology. Subsequently, the pattern was transferred into the silicon layer using Reactive Ion Etching (RIE). Buffered Hydrofluoric Acid (BHF) was then applied to remove the buried oxide layer and release the fabricated beam structures.\\

\textbf{Experimental Setup}.

\begin{figure}[h]
    \centering    
    \includegraphics[width =\linewidth]{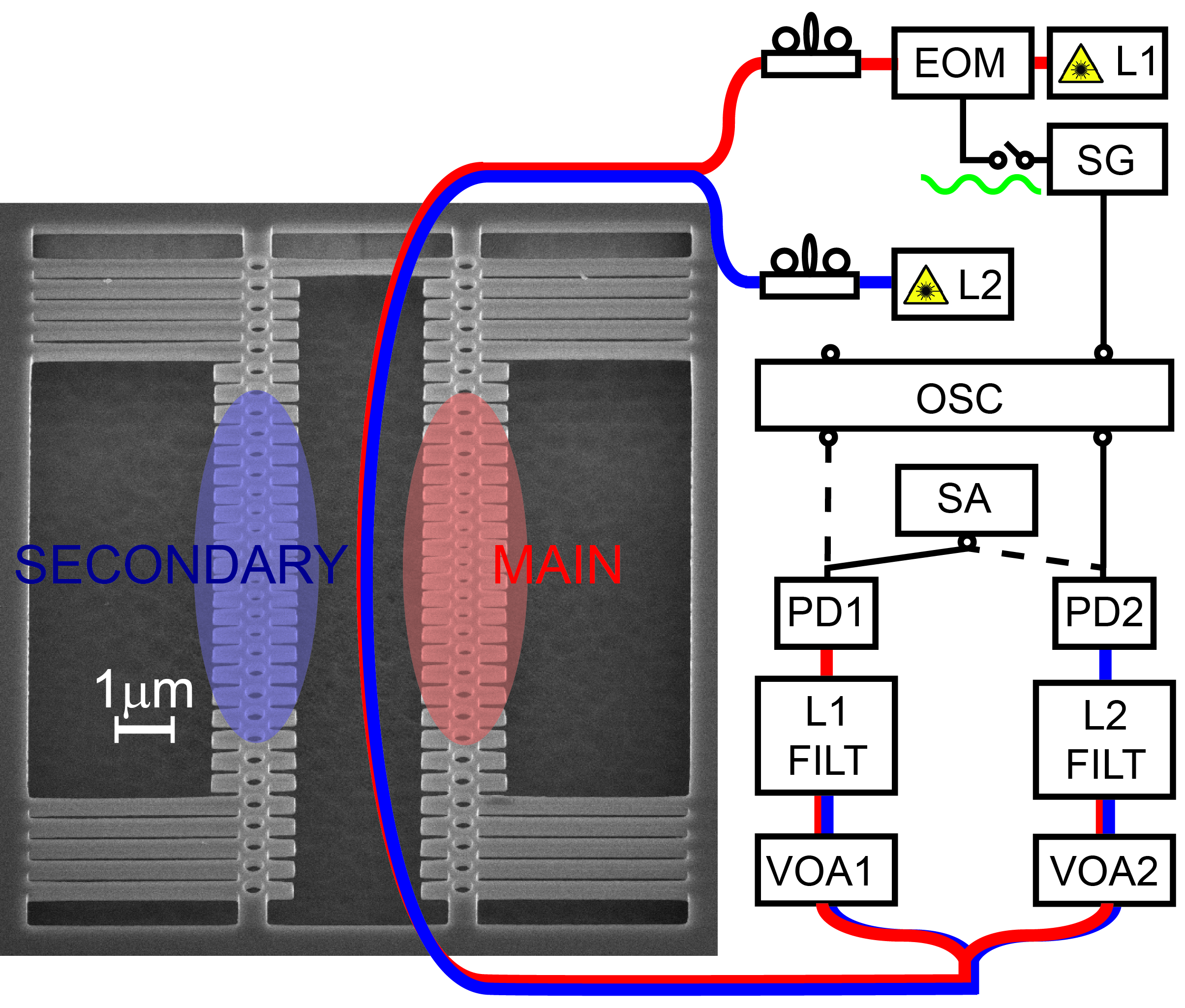}
    \caption{Scheme of the experimental setup. Left panel shows a SEM image of the chip-integrated OM device, where main and secondary OMOs are highlighted in red and blue respectively. Right panel shows an schematic representation of the experimental layout described in the main text. L1 and L2, tunable lasers; EOM, electro-optic modulator; SG, signal generator; OSC, oscilloscope; SA, spectrum analyzer; PD1 and PD2, photodetectors; L1 FILT and L2 FILT, Fabry-Perot wavelength filters; VOA, variable optical attenuator.}
    \label{fig: setup2}
\end{figure}
 
\textbf{Equations of motion of free-carrier, temperature and optomechanical systems}. In this section, we report the mechanism used to drive optomechanical cavities to a coherent, high-amplitude and self-sustained state of mechanical oscillation in the sideband unresolved regime. It is based on the interplay between the thermo-optic (TO) effect and free carrier dispersion (FCD) that emerges in silicon crystals. Both effects have an impact on the refractive index of the material and hence, on the optical resonance of the cavity. The dynamical evolution of free carriers ($N$) and increase in temperature ($\Delta T$) can be expressed as two coupled differential macroscopic equations:

\begin{equation}
\begin{aligned}
 \dot{N} &= - \frac{1}{\tau_{FC}}N + \alpha_{SPA} n_0 (N_0 - N) \\
 &\dot{\Delta T} = - \frac{1}{\tau_{T}}\Delta T + \alpha_{FC}n_0 N \\
 \end{aligned}
\label{eq: system1}
\end{equation}
where the coupling magnitude is the intracavity photons, $n_0 = n_{0,m} \Delta \lambda_0^2/(\Delta \lambda_0^2 + 4 (\lambda_r  - \lambda_l )^2)$, with $n_{0,m} = \frac{2P_l \kappa_e \lambda_0}{\kappa^2 hc}$. Here, $\Delta \lambda_0$ is the linewidth of the  optical resonance at room temperature, $P_l$ and $\lambda_l$ the power and wavelength of the incident laser light, respectively; $\kappa$ the overall damping rate and $\kappa_e$ the extrinsic one. Note that the position of the resonance is modified by the TO and FCD contributions and can be written in first-order approximation as $\lambda_r \approx \lambda_0 - \frac{\partial \lambda_r}{\partial N}N + \frac{\partial \lambda_r}{\partial T} \Delta T$. Regarding the meaning of the coupled system, the first equation takes into account the single-photon absorption (SPA) through $N_0$ intragap states per unit volume and a recombination time of $\tau_{FC}$. The second one considers the fraction of photons that are absorbed and transformed into heat through free carrier absorption (FCA). In that way, $\alpha_{SPA}$  represents the rate of free-carrier density increase per photon and unit of density of available intragap states, while $\alpha_{FC}$ the rate of temperature increase per photon and unit of free-carrier density. It is worth mentioning that the response of $n_0$ to the different contributions its adiabatic since the system is in the regime where $\kappa$ is much larger than the characteristic rates (1/$\tau_{FC}$, 1/$\tau_{SPA}$).\\

 Under specific conditions of the driving laser, the whole system can enter in a dynamical regime described by a self-sustained limit cycle, which we refer as self-pulsing. The modulation on intracavity photons is then transduced linearly to the radiation pressure optical force that drive the oscillator ($F_0 (t) \propto n_0$). Hence, the mechanical modes of the optomechanical crystal can be described as damped linear harmonic oscillators driven by an anharmonic force: 

 \begin{equation}
        \frac{d^2 x(t)}{dt^2} + \Gamma \frac{dx(t)}{dt} + \Omega^2 x(t) = \frac{F_0 (t)}{m_{eff}}
    \label{Eq: mech}
\end{equation}

  The anharmonic modulation of intra-cavity photons induced by the self-pulsing limit cycle at a certain frequency $\nu_{SP}$ can do resonant driving of the mechanical modes, providing amplification and achieving a self-sustained motion of large amplitude (refered as mechanical lasing for the similarity with the optical counterpart). This effect can occur for any of the different M harmonics present in the non-linear modulation if the optomechanical coupling and the input power are sufficiently high (see supplementary material S2). It is worth mentioning that once the mechanical lasing regime is achieved, its contribution to the perturbation generated in $\lambda_r$ through the optomechanical coupling can no longer be neglected:
  \begin{equation}
      \lambda_r \approx \lambda_0 - \frac{\partial \lambda_r}{\partial N}N + \frac{\partial \lambda_r}{\partial T} \Delta T + \frac{\lambda_0 ^2 g_{0}}{2\pi c x_{ZPF}}x
  \end{equation}

\section*{Aknowledgements}
This work was supported by the MICINN projects ALLEGRO (Grants No. PID2021-124618NB-C22 and  PID2021-124618NB-C21) and MOCCASIN-2D (Grant No. TED2021-132040B-C21).

\nocite{*}
\bibliographystyle{naturemag}
\providecommand{\noopsort}[1]{}\providecommand{\singleletter}[1]{#1}%

\section*{S1 One-dimensional optomechanical crystal cavities}

\begin{figure*}[t!]
\centering    
\includegraphics[width= 0.8\linewidth]{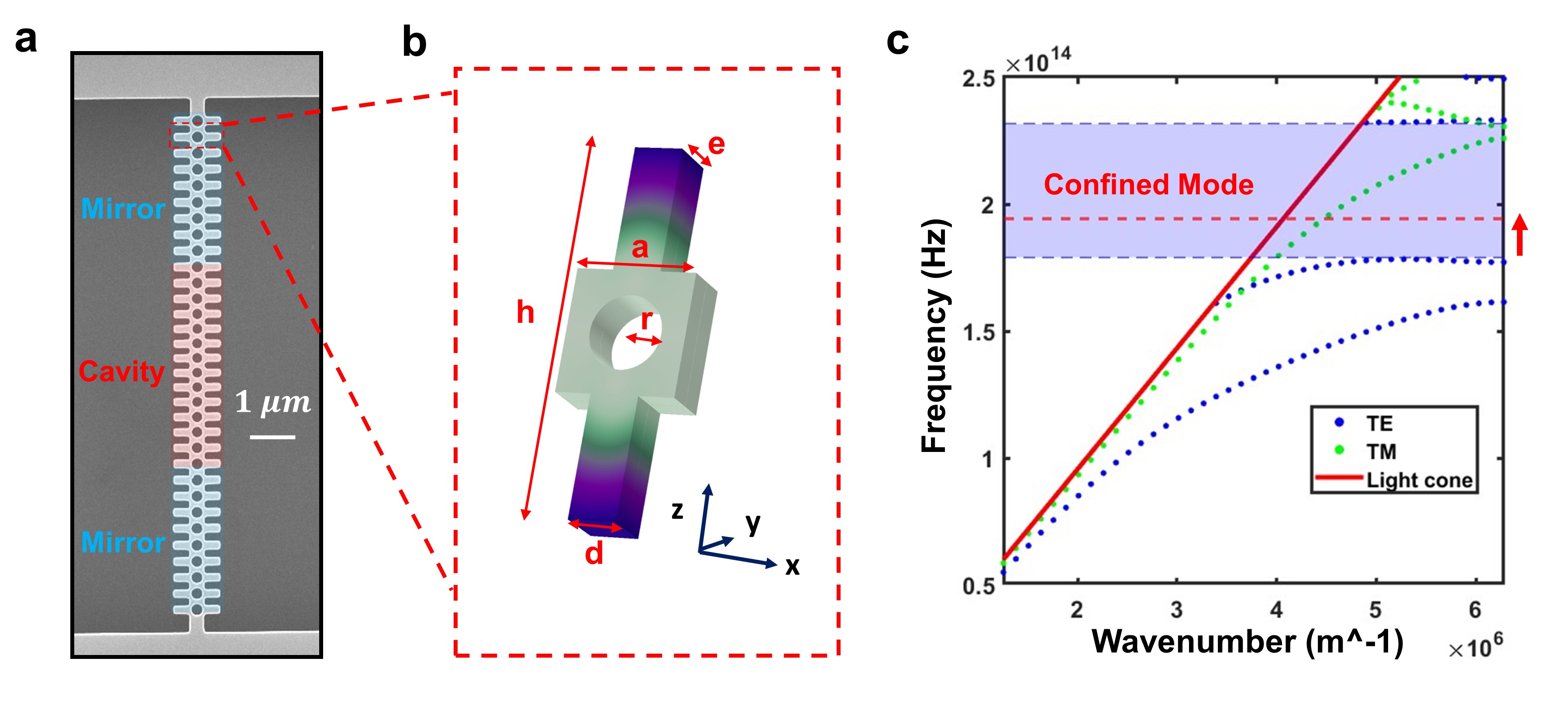}
\caption{Design of the optomechanical (OM) cavity. \textbf{a} SEM image of a fabricated nominal geometry. The cavity region is highlighted in red, where the parameters of the unit cell composing the mirror (blue regions), are adiabatically modified to generate a defect. \textbf{b} Unit cell of the OM crystal with parameters (a,e,r,d,h) = (500, 220, 150, 200, 1500) nm. \textbf{c} Optical band dispersion diagram showing a bandgap for TE modes in the range of 200 THz. Red dashed line represents the frequency of the optical cavity mode, which is constructed from the lower band edge through the adiabatic modification towards the center.  \label{fig: design}}
\end{figure*}

\section*{S2 Numerical simulations of the self-pulsing and mechanical lasing dynamics.}

\begin{figure*}[t!]
    \centering    
    \includegraphics[width= 0.8\linewidth]{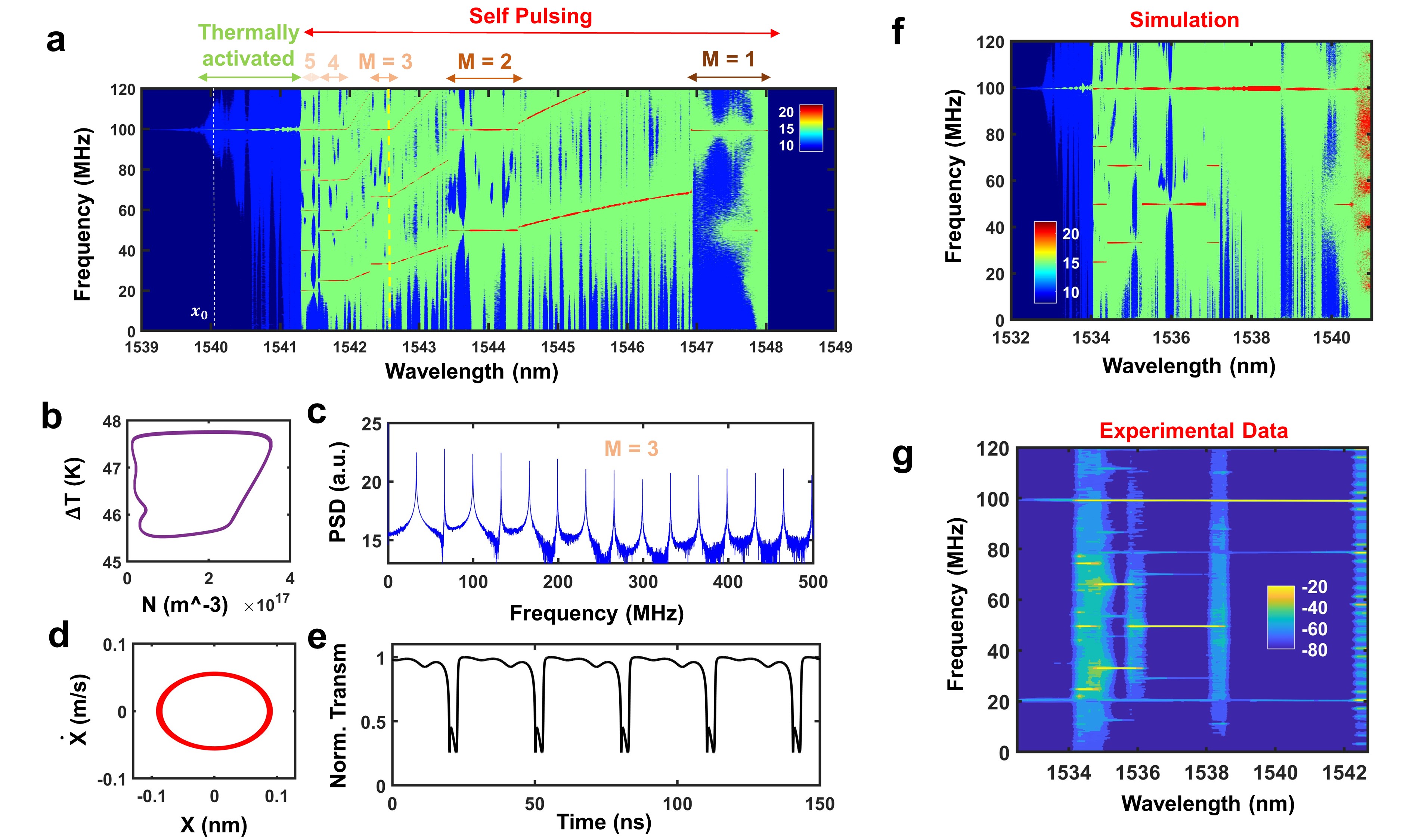}
    \caption{Numerical simulations of the macroscopic model for the self-pulsing mechanism. \textbf{a} RF spectra obtained when performing a Fast Fourier Transform to the computed transmission for different values of the incident laser wavelength. The sweep is done from left (low wavelengths) to right and initial conditions for each iterations are selected as the solution of the previous one. The region where the mechanical modes do not experience amplification is highlighted in green, while the part where the SP is active, in red. \textbf{b-e} Different magnitudes associated to the solution corresponding to the mechanical lasing regime (M = 3), where the third harmonic of the SP is the one providing amplification to the mechanical mode. In particular, the self-limit cycle drawn by free carriers and temperature, RF spectra, mechanical oscillator parameters and computed normalized transmission are shown respectively. \textbf{f-g} Comparison between a simulated contour RF plot (f) obtained in a similar way to the one shown in (\textbf{a}) with experimental data (g). The parameters of the simulation as incident power, damping rate or linewidth are selected to match the ones observed in that measurements.}
    \label{fig: sp}
\end{figure*}

The methods section of the main text presents a model for the description of the anharmonic modulation of the intra-cavity photon number that emerges in silicon due to free carrier dispersion (FCD) and thermo-optic (TO) effects. Here, the system is solved numerically in MATLAB using standard ordinary differential equation (ODE) solving methods. Parameters of the model are extracted from previous works and experimental measurements \cite{freecarrier, temperature, vart, gom}. Initial conditions are settled to [$N$(0), $\Delta T$(0), $x(0)$, $\dot x(0)$] = [$1.5\cdot 10^{17} cm^{-3}$,  $(\lambda_{l} - \lambda_{0})/(\partial \lambda_{r}/\partial T)$, 0, 0]. We select a temporal step of $\Delta t = 2\cdot 10^{-11}s$ to characterize properly the fast dynamics of free carriers and a time span $\Delta t = 1\cdot 10^{-4}s$ for assuring that the system achieves a stationary dynamical regime. Afterwards, a Fast Fourier Transform (FFT) is performed to the temporal trace of the computed transmission (extracted from the intra-cavity photon number) in the stationary solution. This process is repeated while sweeping the wavelength of the incident laser light from lower to higher values. Initial conditions are replaced after the first iteration with the stationary case of previous solutions. In that way, each iteration depends on the previous one, in a similar way to the real experiment. The evolution of the obtained RF spectra is shown in Fig. \ref{fig: sp}a. Initially, when the laser light have just entered resonance (which in this case corresponds to 1540 nm), it is possible to observe a signal oscillating at the frequency of the mechanical mode. Here light is playing a passive role as a probe of the mechanical oscillation, but is not providing amplification of the mechanical motion.  After a certain threshold, the self-pulsing (SP) mechanism is activated, whose harmonics can be visualized as the red curves. When a certain harmonic M resonates with the mechanical mode, it provides amplification and frequency-locks to it (plateaus). Note that the increase of photons in the cavity involves heating, leading to a shift of the optical resonance to longer wavelengths. At some point, the nonlinear system can no longer display a state of self-sustained oscillation and the resonance blue-detune from the laser, going back to its room temperature spectral position.\\

The particular non-linear dynamics at a certain $\lambda_l$ corresponding to the regime of M = 3 is shown in figures \ref{fig: sp} b-e, where the third harmonic of the SP is the one providing amplification to the mechanical mode. Here, the self-limit cycle formed by temperature and free-carrier population is depicted (b), as well as the RF spectra associated with the predicted transmission computed from $n_0$ (c). It is worth noting that in this regime, the amplitude of the mechanical oscillation is around 0.05 nm (d), extending to an order of magnitude higher for the M = 1 plateau. Output transmitted light exhibits a clear nonlinear behaviour (e).  \\

Lastly, in Figs. \ref{fig: sp}g-f we have compared the simulated RF spectrum using a larger power value than that used in Fig. \ref{fig: sp}a with our experimental results, demonstrating significant agreement. In this case, the dynamical solutions of the system do not include isolated SP so that different M mechanical lasing regimes are continuously accessed during the sweep of the laser wavelength, leading to a staircase-like curve.  

\section*{S3 Numerical simulations of mechanically coupled OM oscillators}

 \begin{figure*}[t!]
    \centering    
    \includegraphics[width= 0.8\linewidth]{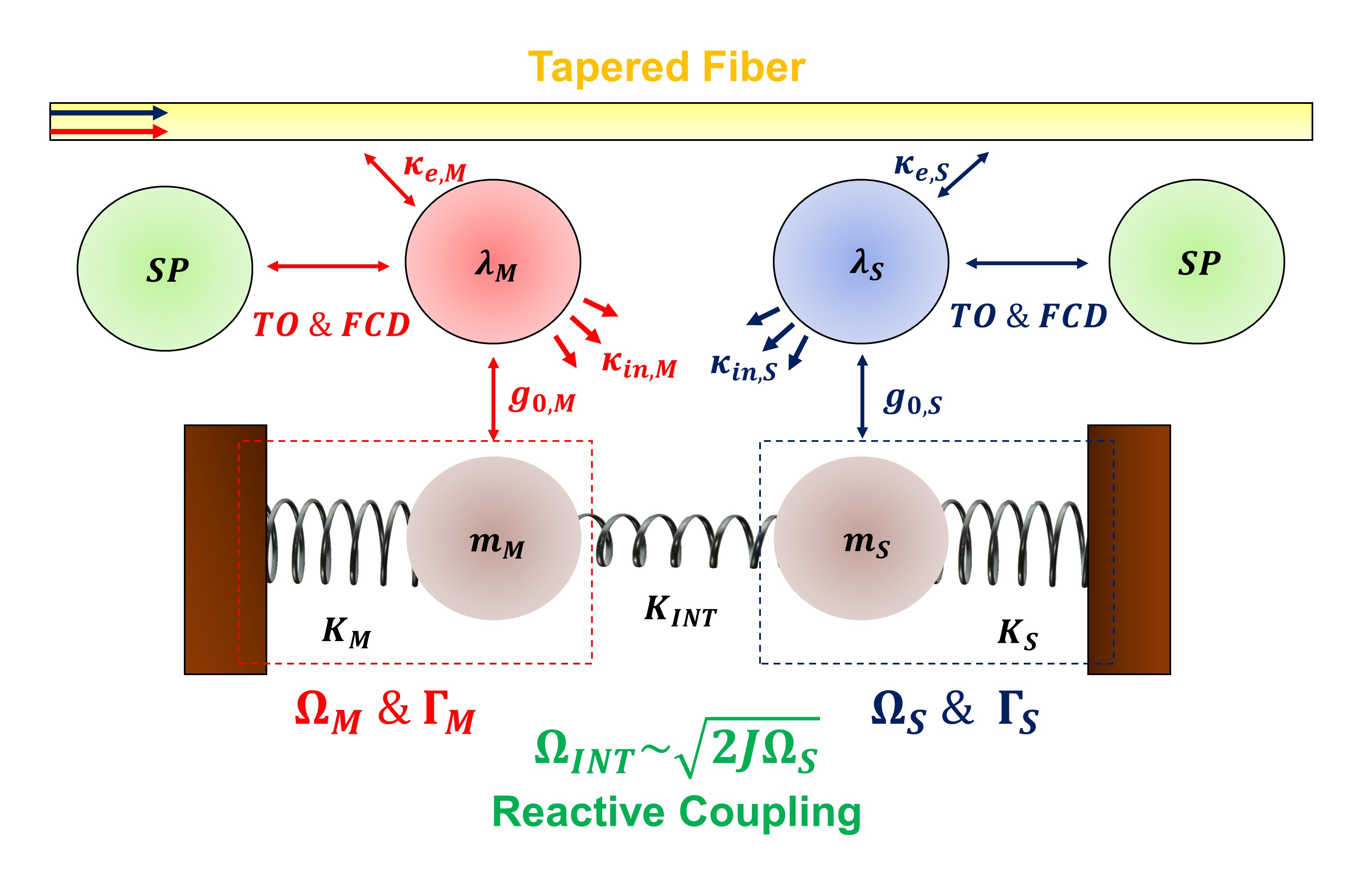}
    \caption{Schematic representation of the OM system. The tapered fiber is highlighted in yellow, while red and blue colors denote each oscillator (main and secondary, respectively). The mechanical part of each OMO is represented as a mass ($m_i$) connected to a spring of elastic constant ($K_i$). Each mechanical subsystem have a certain mechanical angular frequency ($\Omega_i$) and dissipation rate ($\Gamma_i$). The interaction is represented with an spring of lower elastic constant $K_{INT}$ $<<$ $K_M$ \& $K_S$, which result in a coupling rate of $\Omega_{INT} \sim \sqrt{2J\Omega_S}$.}
    \label{fig: scheme}
\end{figure*}

The experiment of the main text considers two mechanically coupled OM crystal oscillators which are self-sustained by the anharmonic mechanism described before. The whole picture can be understood through Figure \ref{fig: scheme}, where the main elements of the system are schematically represented. Two tunable lasers excite both optical modes which support separated optical resonances, having extrinsinc ($\kappa_{e,i}$) and intrinsic ($\kappa_{in,i}$) optical decay rates, respectively. At the same time, the resonant wavelengths are modulated through the TO and FCD effects following the dynamics of the self-pulsing mechanism. The optomechanical coupling communicates the optical and mechanical modes so that this intrinsic modulation of the radiation pressure force can drive the resonators to the mechanical lasing regime (M = 1). Once this regime is achieved, the large amplitude of the mechanical oscillation becomes a non-negligible source of modulation of the intra-cavity photon number. \\

At this point there are two optical channels undergoing a modulation given by the mechanical oscillation of each OM crystal. The mechanical interaction is represented as a spring with low elastic constant ($K_{INT}$ $<<$ $K_{M,S}$) that communicates both oscillators. The coupling is considered to be reactive and can be understood as a restoring force that emerges from the linking tether as it is pulled of its relaxed condition.\\

On the other hand, an external modulation is also applied to the incident power that arrives to the main cavity. Here, the applied voltage to the Mach-Zender electro-optic modulator (EOM) is:  
\begin{equation}
    V(t) = V_{max} sin(2 \pi f_{mod} t) + V_{DC}
\end{equation}
where $f_{mod}$ is the frequency of the external modulation, $V_{DC}$ a voltage that can be applied to operate in the quadrature point. Thus, the power that enters in the main cavity can be read as
\begin{equation}
P_{input} \propto 1 + \cos \left( \pi \frac{sin(2\pi f_{mod}t) + V_{DC}}{V_{\pi}} \right)
\end{equation}
where $V_{\pi}$ is the characteristic voltage to change the phase $\pi$ radians in the EOM. The offset voltage ($V_{DC}$) is set at the quadrature point $V_{DC}$ = 0.5$V_\pi$ to minimize larger harmonics in the perturbation generated by the signal generator (SG) so that, if the maximum voltage of the RF modulation signal is small, the output light power responds linearly. Considering that the intra-cavity photon number is proportional to the incident power, it is arrived to:
\begin{equation}
n_{0}'(t) = n_{0,m} \left(1 - \sin\left(\pi\frac{V_{max}}{V_{\pi}} \sin(2\pi f_{mod}t)\right)\right) n_0(t)
\label{eq: nmod}
\end{equation}
where there self-induced modulation will be enveloped by the external one. \\

\begin{figure*}[t]
    \centering    
    \includegraphics[width= 0.8\linewidth]{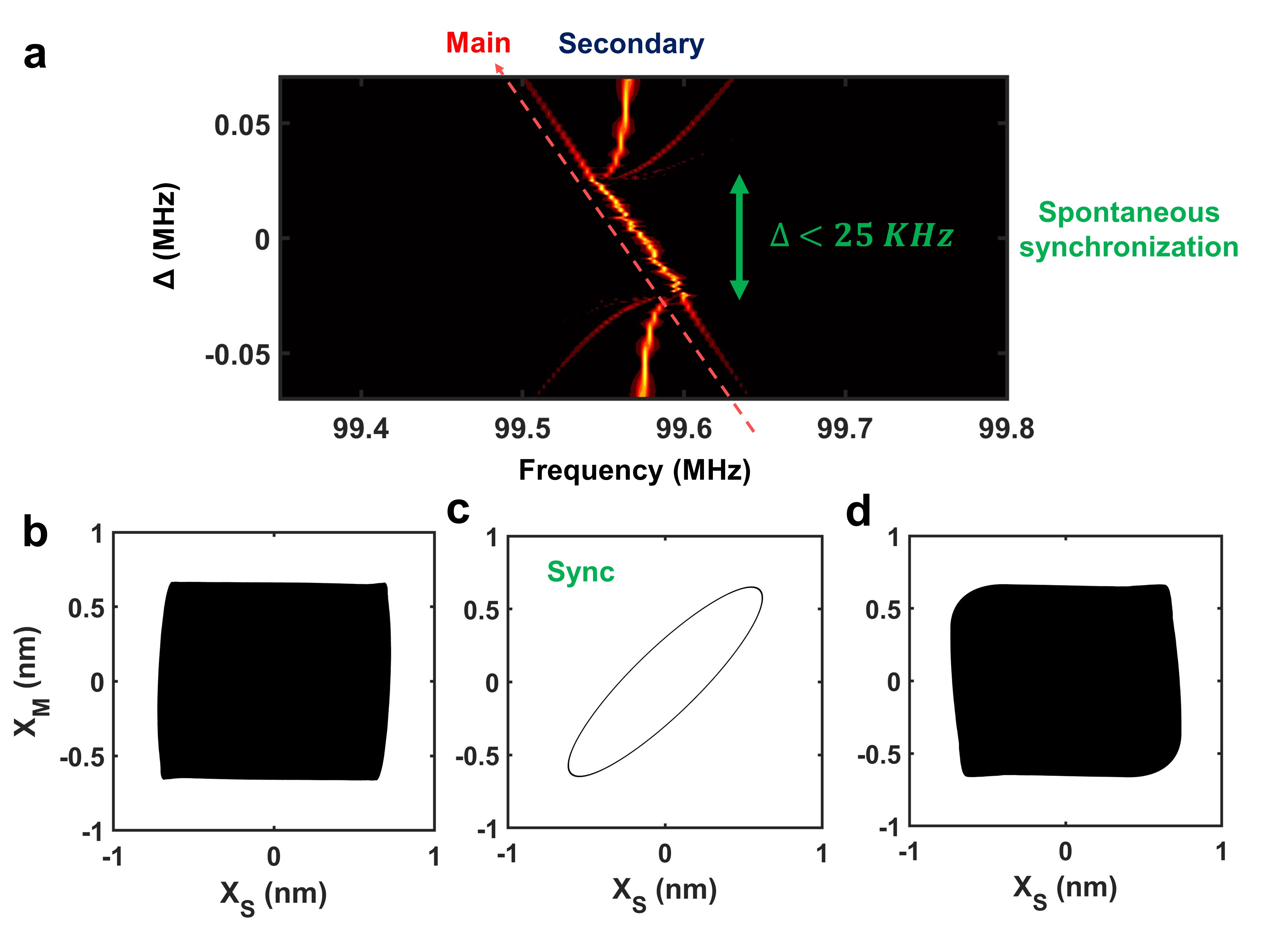}
    \caption{Numerical simulation of the spontaneous synchronization between both OMOs depending on their separation in mechanical natural frequency ($\Delta$). The coupling strength (J/2$\pi$) is set to 18 KHz. \textbf{a} Contour radio-frequency (RF) plot of the normalized Fourier transform of the computed optical transmission in the second oscillator as a function of $\Delta$. The green arrow highlight the range where spontaneous synchronization occurs, while the red dashed one, the sideband associated to the main OMO dynamics. \textbf{b - d} Representations of the phase space covered by the mechanical displacement ($X_i$) of both OMOs for the different regimes observed in the contour plot.}
    \label{fig: analysis1}
\end{figure*}

Considering both, the mechanical interaction and the external modulation in the intracavity photon number, the system of differential equations that describes the dynamics of both oscillators read as:
\begin{align}
 \dot{N_i} &= - \frac{1}{\tau_{FC}}N_i + \alpha_{SPA} n_{0,i} (N_0 - N_i) \label{Eq: N} \tag{8a}\\
 &\dot{\Delta T_i} = - \frac{1}{\tau_{T}}\Delta T_i + \alpha_{FC}n_{0,i} N_i \label{Eq: T} \tag{8b}\\
 \ddot{x_i} + &\Gamma_{i}\dot{x_i} + \Omega_{i}^2 x_i + \delta_{i,S}2J x_M = \frac{\hbar g_{0,i}}{m_{eff,i}X_{ZPF}} n_{0,i} \label{Eq: oscillator} \tag{8c}
 \end{align}
where the sub-index i = M,S denotes main and secondary respectively. As mentioned in the main text, we consider an unidirectional system where the main OMO is oscillating with much larger mechanical amplitude. Hence, the interaction is only present in the secondary OMO dynamics since the contribution of $x_s$ has been neglected 2J($x_M$ - $x_S$) $\approx$ 2J$x_M$.  Regarding the intracavity photons, the first oscillator is the one that receives the external modulation, thus $n_{0,M}$ corresponds to the modulated case (Eq. \ref{eq: nmod}) while the temporal evolution of $n_{0,S} = n_0$ is only governed by the modulation on the optical resonance of the cavity due to the TO and FCD effects as well as the mechanical motion. This means that the secondary oscillator is only aware of the external modulation through the mechanical coupling term. The intrinsic magnitudes of the model (dissipation rates, intragap-states...) have been considered the same for both oscillators.\\

Even if Eqs. \ref{Eq: N}, \ref{Eq: T} and \ref{Eq: oscillator} represent a system of nonlinear coupled differential equations and do not have analytical solution, it is possible to solve it numerically as a problem of eight first order differential equations. First, it is necessary to set the incident laser wavelength so that both oscillators are in the mechanical lasing regime in the absence of mechanical coupling and external modulation. To do that, parameters J and $V_{max}$ are settled to 0 and an analysis is performed to find the properly detuning, which is chosen to be $\lambda_{l,i} - \lambda_{0,i} = 4.5$ nm. Given that in the experiment both mechanical oscillators are relatively close in frequency (around tens of KHz), simulations require a high-resolution to be able to resolve that close-by peaks. This can be achieved by increasing the time span to 4e-4, which provides a resolution of 3KHz after the Fourier transform. \\

\underline{Spontaneous synchronization}\\

\begin{figure*}[t]
    \centering    
    \includegraphics[width= 0.9\linewidth]{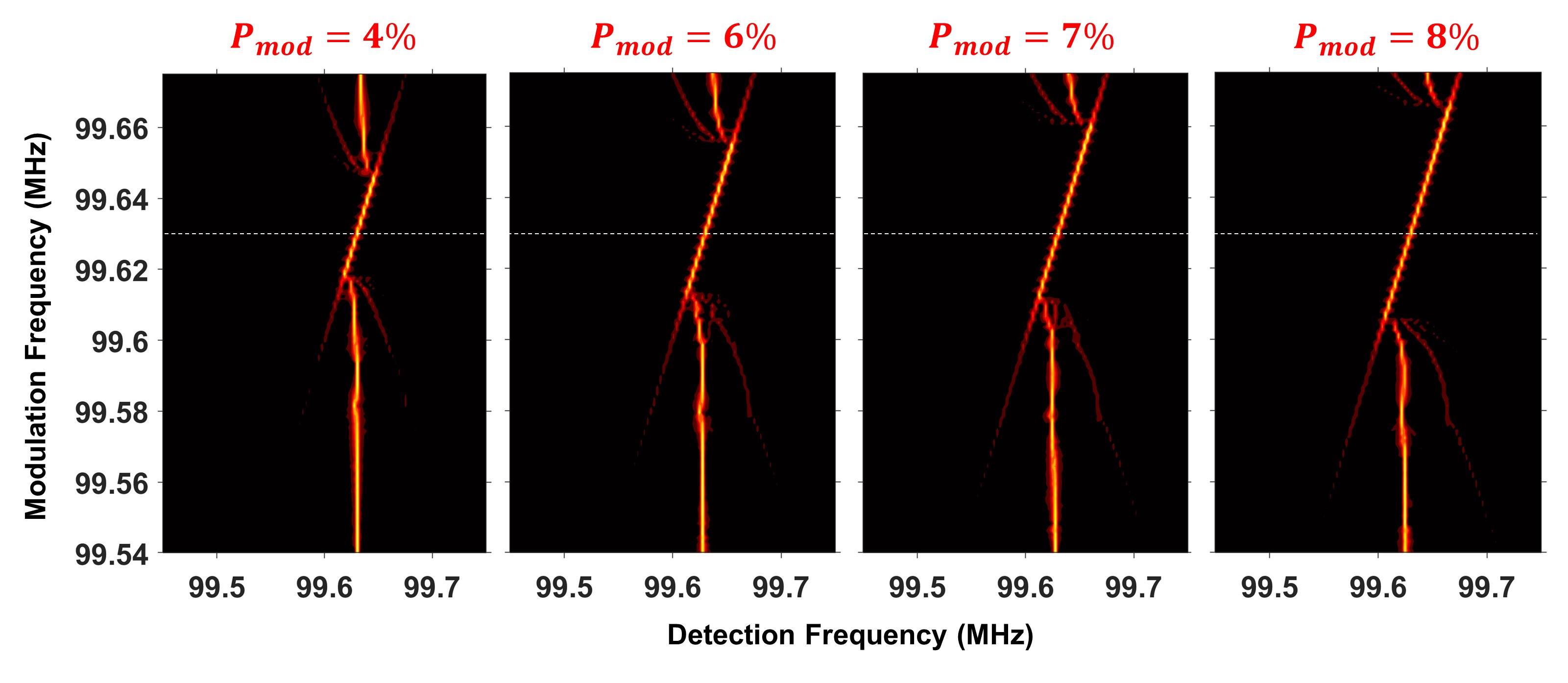}
    \caption{Numerical simulations of the injection locking of the external modulation signal to the main OMO dynamics. Each panel shows the normalized fast Fourier transform of the optical transmission in the main cavity when performing a sweep in the modulation frequency from lower to higher values. The selected amplitude of modulation is highlighted above each contour plot. White dashed line indicates the starting value for the locking regime in the case of low power modulation.}
    \label{fig: analysis2}
\end{figure*}

This section deals with the case in which no external modulation is applied. Once both oscillators are in the mechanical lasing regime, it is possible to perform a similar analysis to the first experiment of the main work (Figure 1e). We have varied the reactive coupling strength to select the proper value that reproduces the spontaneous synchronization range of the experiment. Fig. \ref{fig: analysis1}a shows the contour RF plot of the computed spectra for the secondary OMO intra-cavity photons dynamics, when varying the mechanical frequency of the main OMO and hence, the detuning between both oscillators. The main OMO dynamics is observed as sidebands that approach to the center peak as the mechanical detuning between the frequencies both resonators decreases. At the beginning, since both oscillator are not synchronized, their mechanical displacements cover most of the phase space when plotted against each other (Fig. \ref{fig: analysis1}b). When their natural frequencies are close enough, spontaneous synchronization occur and their displacements follow a closed trajectory (Fig. \ref{fig: analysis1}c). Finally, when the system exits the Arnold tongue, and the mechanical displacements plotted in the phase space fill the graph again (Fig. \ref{fig: analysis1}d).  \\

Here J/2$\pi$ is set to 18 KHz, providing a spontaneous synchronization range of 50 KHz, similar to that observed in the experiment. It is important to mention that in the simulation, the unidirectional character of the interaction is introduced just by including the interaction term on the secondary OMO equation and neglecting $x_S$ with respect to $x_M$. However, in contrast to the real experiment, both oscillators exhibit similar amplitudes. Since the spontaneous synchronization range depends on the amplitude of the mechanical perturbation that the secondary OMO is receiving, the amplitude of the mechanical oscillation of the main OMO plays an important role. Hence, the actual interaction strength could be one order of magnitude lower, confirming the weak character of the coupling and indicating that fabrication disorders are the main source of splitting between the mechanical resonant frequencies.
\\

\underline{Cascaded injection locking}
\\

In this section we include the external modulation with a certain power  $P_{mod} = 100 Sin(\pi V_{max}/V_{\pi})$, aiming to simulate the cascade injection locking experiment. In the simulations we establish a frequency difference of 45 KHz between both OMOs, thus assuring no spontaneous synchronization occurs in the absence of an external input. Figure \ref{fig: analysis2} focuses in the main oscillator (the one that directly receives the modulated signal), where it is represented the fast Fourier transform of transmitted light. Spectra are represented in a normalized linear scale with arbitrary units. Here, a sweep in $f_{mod}$ is done from low to high frequencies to observe the range of injection locking depending on the power of modulation. The step taken for performing the sweep in external modulation frequency is set to 1 KHz.\\

 \begin{figure*}
    \centering    
    \includegraphics[width= 0.9\linewidth]{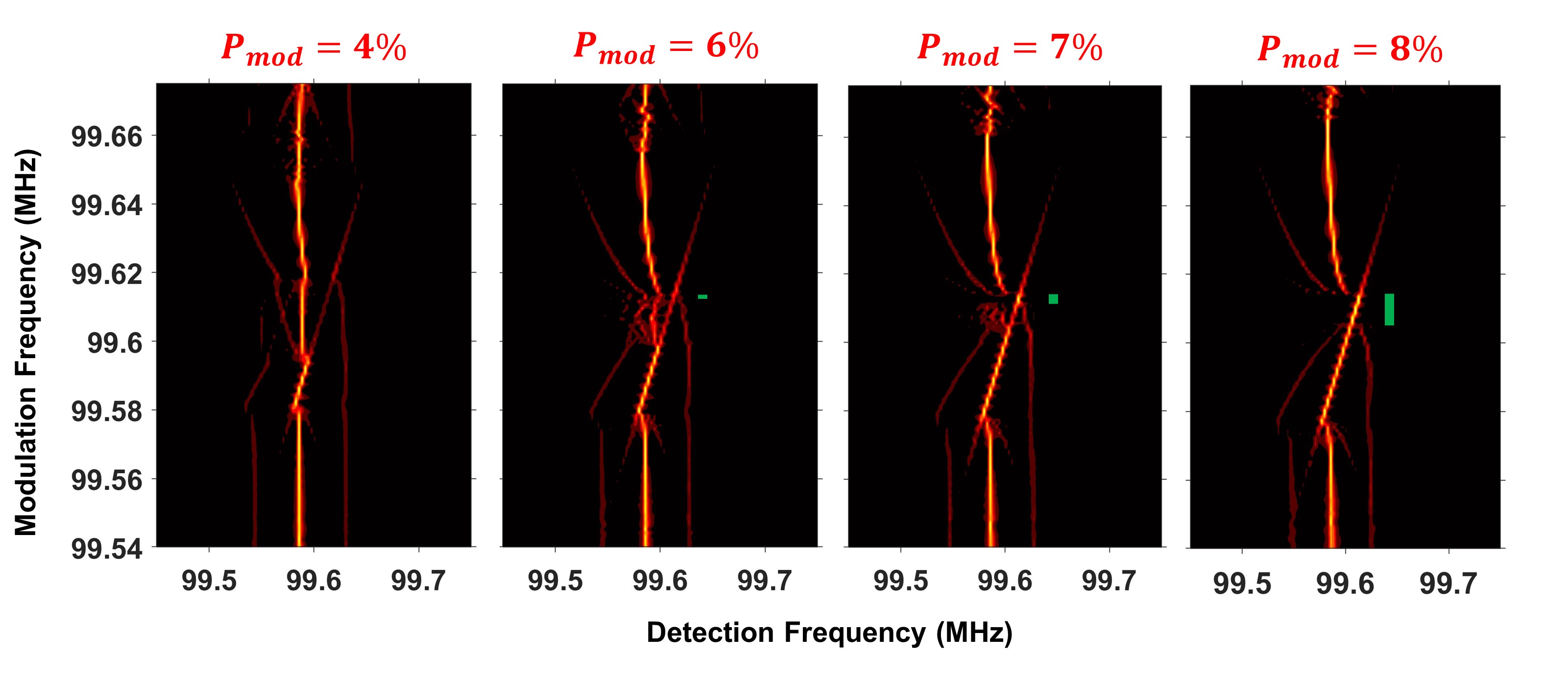}
    \caption{Countour RF plots in a normalized linear scale of the fast Fourier transform of the optical transmission in the second cavity when performing a sweep on the external modulation frequency. Panels from left to right have been computed using different amplitudes of external modulation.  The most intense RF signal corresponds to the secondary OMO natural frequency, while the sidebands corresponds to the transduced motion from the main oscillator due to mechanical coupling. The CIL regime is highlighted with a green thick line.}
    \label{fig: analysis3}
\end{figure*}

Regarding the range for injection locking, we observe an increasing on its value with $P_{mod}$ as expected from previous simulations \cite{injectionlocking}. The locking range expands particularly in the region after the initial locking (white dashed line), which corresponds to the range where the frequency of external modulation is above the natural one of the oscillator. Finally, it is worth to mention that the natural frequency of the second oscillator is not present in these graphs since the coupling term has not been added to the system of the main oscillator (Eq \ref{Eq: oscillator}).\\

The next step is to analyze the transmission of the second oscillator (the one that receives the modulation indirectly through the mechanical link). Figure \ref{fig: analysis3} shows the results of this analysis As we expected, the mechanical oscillation of the main OMO is now observed in the secondary oscillator dynamics. Even if this modulation arrives to it in an indirect manner, the mechanical perturbation is enough to generate injection locking of the secondary OMO dynamics to the external signal, reproducing the experimental observations. Now, as we increase the modulation power, the locking ranges increase and start to appear a region where both, the main and the secondary oscillators are locked to the external modulation, what we call cascade injection locking (CIL). Simulations also predict the expansion of this range when increasing the modulation amplitude. \\

\end{document}